# A Pilot Study on
# Coupling CT and MRI through Use of Semiconductor Nanoparticles

Matthew Getzin, Lars Gjesteby, Yen-Jun Chuang, Scott McCallum, Wenxiang Cong,
Chao Wang, Zhengwei Pan, Guohao Dai and Ge Wang*
Biomedical Imaging Center/Cluster
Center for Biotechnology and Interdisciplinary Studies
Department of Biomedical Engineering
Rensselaer Polytechnic Institute, Troy, NY 12180, USA
*ge-wang@ieee.org

**Abstract**

CT and MRI are the two most widely used imaging modalities in healthcare, each with its own merits and drawbacks. Combining these techniques in one machine could provide unprecedented resolution and sensitivity in a single scan, and serve as an ideal platform to explore physical coupling of x-ray excitation and magnetic resonance. Molecular probes such as functionalized nanophosphors present an opportunity to demonstrate a synergy between these modalities. However, a simultaneous CT-MRI scanner does not exist at this moment. As a pilot study, here we propose a mechanism in which water solutions containing $LiGa_5O_8$:$Cr^{3+}$ nanophosphors can be excited with x-rays to store energy, and these excited particles may subsequently influence the $T_2$ relaxation times of the solutions so that a difference in $T_2$ can be measured by MRI before and after x-ray excitation. The trends seen in our study suggest that a measurable effect may exist from x-ray excitation of the nanophosphors. However, there are several experimental conditions that hinder the clarity of the results to be statistically significant up to a commonly accepted level ($p=0.05$), including insoluble nanoparticles and inter-scan variability. Nevertheless, the initial results from our experiments seem a consistent and inspiring story that x-rays modify MRI $T_2$ values around nanophosphors. Upon availability of soluble nanophosphors, we will repeat our experiments to confirm these observations.

Acronyms: Nanophosphors (NPs), computed tomography (CT), magnetic resonance imaging (MRI), echo time (TE), repetition time (TR)

## 1. Introduction

A major challenge in medical imaging is to achieve both fine resolution and high sensitivity in a single scan for diagnostic and interventional purposes. Multi-multimodal imaging hardware, such as PET-CT and PET-MRI, represents a great step toward meeting this challenge, but synergistic integration of CT and MRI into a single machine has yet to be realized. One strategy for demonstrating a coupling relationship between CT and MRI is through *in vivo* or *in situ* use of doped semiconductor nanoparticles, which can be functionalized as molecular probes to enable high biological specificity.

Nanoparticles have been used *in vivo* and *in situ* to provide contrast enhancement in CT and other imaging techniques [1–3] and deliver therapies as well. Recent studies have shown that $LiGa_5O_8:Cr^{3+}$ nanophosphors irradiated with x-rays or ultraviolet light can exhibit persistent luminescence for up to 1,000 hours after excitation [4]. The excitation energy in this process is stored in the nanophosphors by pumping electrons into the energy trap, which can be released by subsequent optical stimulations [4]. Nanophosphors which persistently hold electrons in the energy trap state are most useful for our investigation on possible MRI readout of x-ray excitation in deep tissues.

Over the past several years, some interesting imaging technologies have been developed that integrate persistent luminescent nanophosphors into established imaging methods. One such technique is stored luminescence computed tomography (SLCT), which aims to "localize and quantify a distribution of energy-storing nanophosphors" with x-ray excitation [5]. Another technique, x-ray micro-modulated luminescence tomography (XMLT), was proposed that enables super-resolution biological investigations with nanophosphors by coupling focused x-ray and luminescence emission [6].

In this initial report, we explore a mechanism in which the $T_2$ relaxation times of aqueous solutions measured by MRI can be modulated by x-ray excitation through interactions of water with nanophosphors such as $LiGa_5O_8:Cr^{3+}$. The x-ray excited nanophosphors simultaneously enable high-resolution imaging, biological targeting, and enhanced MRI soft tissue contrast *in vivo* or *in situ*. In our explorative trial, we use aqueous solutions of nanophosphors under the hypothesis that x-ray excited electrons in the nanophosphors can change the local magnetic field of the solution in a measurable way, which specifically leads to measurable changes in MRI relaxation parameters of the solutions before and after x-ray excitation. This technique, referred to as nanoparticle-enabled x-ray MRI (NXMRI), could synergistically blend merits of optical imaging, CT, and MRI in an unprecedented manner. Imaging hardware capable of simultaneous CT-MRI imaging, which has yet to be prototyped, would provide an ideal platform for our proposed multi-physics coupling.

## 2. Materials and Methods

All the nanophosphors used in the following experiments were $LiGa_5O_8:Cr^{3+}$ as described by Chuang et al. [4,7]. The size of the sample particles used in the experiments was indicated to be <100 nm. These particles are highly non-colloidal, settling out of an aqueous solution within 2 minutes. Long-term settling (> 1 hour) resulted in hypointense regions within the MRI images indicative of a lack of hydrogen molecules to give readable

MRI signals (images not shown).  For this reason, the samples had to be agitated before each set of acquisitions and excitations.  This resulted in a lack of control over the particle distribution within any given sample between before- and after-excitations.

### 2.1. Slurry Phantom Creation

The measured phosphor content in each sample through bulk concentration calculations was deemed as misrepresentative of the true local concentrations that resulted from the quick settling behavior of the nanophosphors when placed in water.  To resolve this problem, we incrementally added small amounts of "dispersed" nanophosphors in water to a 3 mm diameter capillary tube.  After the particles visibly settled, the supernatant water was removed from the sample and more "dispersed" nanophosphors were added to the sample.  This stepwise addition of nanophosphors continued until the capillary tube had approximately a 7 mm tall column of nanophosphors. After the required volume was achieved, the remaining supernatant was kept on top of the net volume of nanophosphors to be used as the proton source for all MRI measurements.

Two samples were made for each trial, *i.e.* x-ray excited, UV excited, unexcited nanophosphors. Please note that the unexcited sample was not subjected to any x-ray excitation during the experiments. Hereafter we refer to scans as either "before-excitation" and "after-excitation" to distinguish the two time points of the scans even though not all samples in the "after-excitation" scan underwent excitation. Additionally, a capillary tube was filled with water to serve as a reference.  The seven capillary tubes were then arranged into a hexagonal shape, wrapped into a bundle with parafilm, and placed in a small polystyrene tube surrounded by $Cu_2SO_4$-doped water.  Figure 1A and B show sagittal and axial views of the phantom arrangement in the polystyrene holder. Furthermore, Figure 1C shows the specific placement of the samples as placed in the MRI imager.

### 2.2. X-ray Irradiation

The x-ray excited samples were prepared in a Scanco vivaCT 40 and scanned at a tube voltage of 70 kVp and tube current of 114 µA.  The pitch and number of projections were set to have the scan run for at least 15 minutes according to the imaging software.

### 2.3. UV Irradiation

The UV excited samples were prepared on the bottom of a Spectrolinker XL-1000 device which produces UV radiation at 254 nm, 2 A, and 120 V.  The samples were irradiated for 15 minutes.

### 2.4. Excitation Verification

To verify the persistent luminescence from the nanophosphors, preliminary observations were made by Nikon TE2000 wide-field microscopy (inverted epi-fluorescence mode; 4X objective lens) with an emission filter (bandwidth 60 nm at 720 nm, Chroma). The microscope was housed in a transparent plastic chamber, in which a "dark-room" was created by black-cloth coverage. For homogenous close-up stimulation, the continuous-wave laser stimulation source at 635 nm (single-mode semiconductor CW laser, S1FC635, Thorlabs) was coupled to a fiber cable (OceanOptics) and a megapixel ultra-low-distortion compound lens (2/3", 25 mm, C-mount, M2518-MPW2, Computer). The laser stimulation power was set at 2 mW for each exposure time of 5 seconds.

### 2.5. MRI Relaxometry Acquisition

The respective phantoms were placed in a Bruker 13 cm, 7T horizontal bore MRI scanner, centered in a 23 mm RF coil. Special care was taken to preclude air bubbles that were present in the regions to be imaged. Prior to the $T_2$ experiments, the position of each phantom was optimized and established using a gradient echo TriPilot scan. The local magnetic field about the field of view (FOV) was further optimized using the FieldMap routine. The $T_2$ experiments consisted of a multi-spin, multi-echo protocol in which the TEs were varied as a train of 16 equally spaced echoes of 10.5 msec with a TR of 2000 msec. The FOV was 20 mm X 20 mm with an in-plane resolution of 0.156 mm X 0.156 mm and slice thickness of 1.25 mm. Six slices were taken to collect as many usable image data as possible. These parameters led to a scan time of 2 minutes and 13 seconds. The resulting dataset from a $T_2$ experiment was processed as a 4-D matrix with dimension of 128 X 128 X 6 X 16 (x, y, z, TE) whose voxel values represent signal intensity modeled by Eq. 1:

$$S = k\rho \left(1 - \exp\left(-\frac{TR}{T_1}\right)\right) \exp\left(-\frac{TE}{T_2}\right), \tag{1}$$

where S is the amplitude of the signal, *k* is a proportionality constant that depends on the machine and RF coil, the unknown $T_1$, $T_2$, and *ρ* (proton density) can be imaged at will.

### 2.6. Relaxation Mapping

The 4-D dataset was pixel-wise fit into Eq. 2 across the TE dimension using the Image Sequence Analysis tool of the ParaVision software (v5.1) as supplied by Bruker:

$$S = A \exp\left(-\frac{TE}{T_2}\right) + c. \tag{2}$$

Eq. 2 is a simplified version of Eq. 1 with *A* being the machine specific proportionality constant *k* multiplied by the sample specific constant *ρ* (proton density) and the (1 – exp(-TR/$T_1$)) function, since TR is held constant in the experiment. The resulting dataset from this analysis step is a 3-D matrix with dimensions of 128 X 128 X 6, and will be henceforth referred to as the $T_2$ map. This dataset was further reduced by manual selection of the slice that contained the largest number of usable samples. Samples were considered usable when a clear interface could be seen between the nanophosphor and water regions.  Because of this restriction, one UV excited sample and one unexcited sample were excluded from the analysis because none of the slices contained a usable interface.  An additional output from the ParaVision $T_2$ mapping software is the intensity image that was also used as the proton-density-weighted image, although it is more representative of the image of the raw intensities at the first TE.

### 2.7. Semi-automatic Edge Analysis

Taking into account the nanophosphor settling as well as the inner- and outer-sphere relaxation mechanisms of contrast agents modeled by Solomon-Bloembergen-Morgan (SBM) relaxation theory [8,9], the change in $T_2$ relaxation is hypothesized to be found at the interface between the nanophosphor and water in the samples.  Analysis of the interface was conducted using common image processing and segmentation algorithms including interpolation, Otsu thresholding, and Sobel edge detection.  Some manual intervention was also needed to select ROIs and choose the edges found to be representative of the nanophosphor-water interface.  All of these techniques were implemented in a customized MATLAB code which can be found in the supplemental material.

The Otsu thresholding and Sobel edge detection were performed on the proton-density-weighted ROIs.  This outcome was treated as the "true" interface between nanophosphor deposit and water.  Once the edges were identified, they were mapped to the $T_2$ image, and the values from the pixels along the edge were averaged.  To reduce the measurement error, edges were manually restricted to the areas of qualitative contrast between nanophosphors and water.  Figure 2 shows the workflow of data analysis.  This same pipeline was repeated four more times for each sample adjusting the Otsu threshold value by 2 grayscale units in both the water (Threshold – value) and NP (Threshold + value) directions to test the robustness of the data trends.

## 3. Results

### 3.1. Excitation Verification

Figure 3A and B show the microscopy images with laser stimulation off and on, respectively. The laser stimulation on-off observations were repeated several times with the NPs sample intentionally exposed to the room-light illumination between each trial. It could be safely concluded that the persistent luminescence is subject to negligible leakage under room-light illumination while the CW laser illumination could effectively stimulate the release of the stored UV energy for narrowband luminescence re-emissions at 716 nm.

### 3.2. Edge Analysis

Figures 4-7 show a comparison of mean $T_2$ time constants along the water-NP interface before- and after-excitation.  From these figures, an effect from excitation is suggested.  In each of the excited samples (UV – Fig. 5, X-ray1 – Fig. 6, X-ray2 – Fig. 7) a decrease in mean $T_2$ is possibly evident although the standard deviations overlap.  The unexcited control (Fig. 4), however, shows a slight but insignificant increase in mean $T_2$ along the edge as seen in Figure 4.  With this control in mind, excitation of the nanoparticle slurries via UV and X-ray seem to show a trend toward a decreased $T_2$ after excitation relative to before excitation.

Further mean comparisons were made with adjusted threshold values.  In this analysis, the edges were calculated after implementing adjusted threshold values and calculating the mean $T_2$s along the new edges.  Figures 8 and 9 show the trend of $T_2$ increase when edges were taken more toward the water region (Thresh – value) and $T_2$ decrease when edges were take more toward the NP region (Thresh + value).

Direct comparison for each of these edges was made to see if a change could be consistently detected between before- and after-excitation images as the edge was shifted into either of the regions.  Figure 10 shows that in the excited samples (UV – B, X-ray1 – C, X-ray2 – D) the mean differences between the before- and after-excitation samples were increased as the edge moves further into the water regions.  The unexcited control (Fig. 10A) also shows an increase in the mean difference between before- and after-excitation into the water region, but the trend for the unexcited control is not as consistent and distinct as for the excited counterparts.

## 4. Discussions and Conclusion

The above-reported data have shown an encouraging trend in which the water coupled to nanophosphor slurries and excited by ionizing radiation such as UV or X-ray for 15 minutes is detected as having reduced $T_2$ relaxation time constants. Changes in relaxometry were most clearly observed along the edge of the slurry-water interface with changes extending somewhat into the bulk water phase.  This trend is consistent with the physical understanding of the redistribution of electrons within the semi-conductor nanophosphors and the subsequent effects on nuclear relaxation.  However, the acquisition and analysis of the data presented here are complicated by the non-colloidal behavior of the nanoparticles employed.  The following remarks are an attempt to disclose and address the issues that result from the non-colloidal behavior while at the same time offering solutions for future efforts.

First, the nanophosphors used in this study were chosen for their unique ability to remain in a stable x-ray excited state over long periods of time.  This property allowed for the particles to be excited outside of the MRI machine and remain excited during the high-resolution imaging sessions.  Despite this vital advantage, these nanophosphors behaved quite poorly in aqueous environments.  Within minutes of dispersing the particles in the water > 90% of the particles had settled to the bottom of the capillary tubes.  Furthermore, after allowing the particles to settle for about an hour, water being excluded from the particle slurry was seen by a loss of MRI readable signal (not shown) as a function of time.

This non-colloidal behavior led to multiple issues related to sample control, stability, and homogeneity. While a robust characterization of a colloidal suspension would have been preferred and more straightforward, qualitative changes in water relaxation about the nanoparticle slurry-bulk water interface were detected once the samples were allowed to settle for 5-10 min before undergoing MRI scans. However, variations in the water-slurry interfaces among the samples and scans of the same sample pre- and post-x-ray treatment surely resulted in measurement variations. Further measurement error, possibly resulting in lower measured changes or larger standard deviations, may have also resulted from variable NP concentrations in the water regions of each sample. Regardless of the source of the variation, overlapping standard deviation ranges between samples before- and after-excitation indicates compromised statistical significance.

Nearly all of these experimental complications can be mitigated through the use of semiconducting nanophosphors that can be more easily suspended or remain colloidal in aqueous solutions. Techniques for improving the water compatibility include milling or crystal growth to smaller particle sizes, coating or conjugating the nanophosphors to hydrophilic polymers. These techniques also open doors for specific targeting of these nanophosphors to cells and proteins. However, such conjugation may also decrease the direct interaction of water with the nanophosphor's electron distribution resulting in less change in $T_2$ after UV or x-ray excitation. Studies using such nanophosphors are already being planned with our collaborators. With enhanced experimental control, robust statistical analysis methods such as paired t-tests can be used to help quantitatively demonstrate the effect of UV or x-ray excitation of nanophosphors on the $T_2$ relaxation of surrounding water protons in a biologically relevant model.

Additionally, uncertainties in the $T_2$ constants for water were high due to our highly-efficient but sub-optimal sampling of its intensity during the relaxation period. In our study, the max TE was ~180 msec. Therefore, trying to fit water $T_2$ constants that are greater than 300 msec likely resulted in some errors that can be improved with a better imaging protocol. In future experiments with two primary $T_2$s, a second $T_2$ experiment with better TE sampling may be necessary for better $T_2$ fits with less variation from voxel to voxel in both types of ROIs.

Because of the low resolution of the images at the sample level, each image underwent bicubic interpolation. Although such methods are accepted for approximating data among measured points, the various thresholds and subsequent edges calculated in our study have also resulted in approximation errors during the $T_2$ averaging.

Finally, to make sure if the change in $T_2$s shown in the data is a result of the nanophosphor excitation, a repeated measurement with the same samples after discharging the nanophosphors could have been done. Unfortunately, the persistence shown by these particles makes true discharging of the nanophosphors a rather challenging task.

In conclusion, despite the list of critiques that have been just made to cast doubt over the source of the $T_2$ changes seen in the experimental data, there appears to be consistent evidence that is physically appealing, and backs the claim that x-rays can be used to alter measured MRI signals through the use of semi-conducting nanophosphors. At the very least, these results will serve as the motivation for further research into this fascinating topic that has clearly a tremendous biomedical relevance. As soon as water soluble nanophosphors become available to our group, extreme care will be taken when repeating and improving our imaging protocols so that the issues discussed above are well addressed. With the rigorous control over the samples, particle distributions, echo time sampling, etc., we believe that the link between x-rays, nanophosphors, and MRI can be utilized to advance imaging technology and deepen the understanding of multi-physics coupling.

## 5. Acknowledgements

We would like to thank Taylor Dorsey for her help in obtaining the excitation verification data.

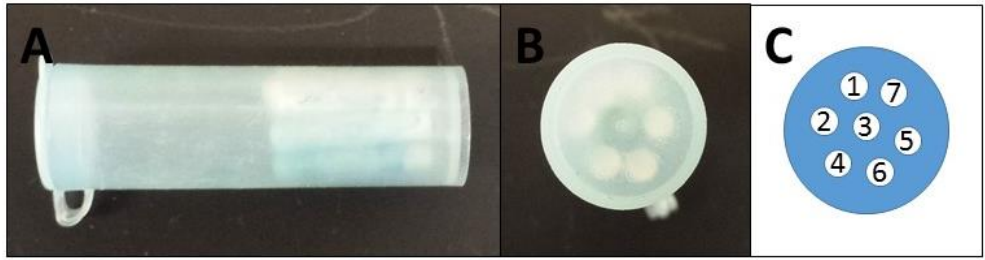

Figure 1. (A) Sagittal view and (B) axial view of the capillary tubes arranged in the phantom and placed in a plastic tube (no $Cu_2SO_4$-doped water is shown in these images). (C) The arrangement of the capillary tubes for the $T_2$ experiment was as follows: 1 and 2 were UV-excited NPs, 3 and 4 were unexcited NPs, 5 and 6 were X-ray excited NPs, and 7 was deionized water.

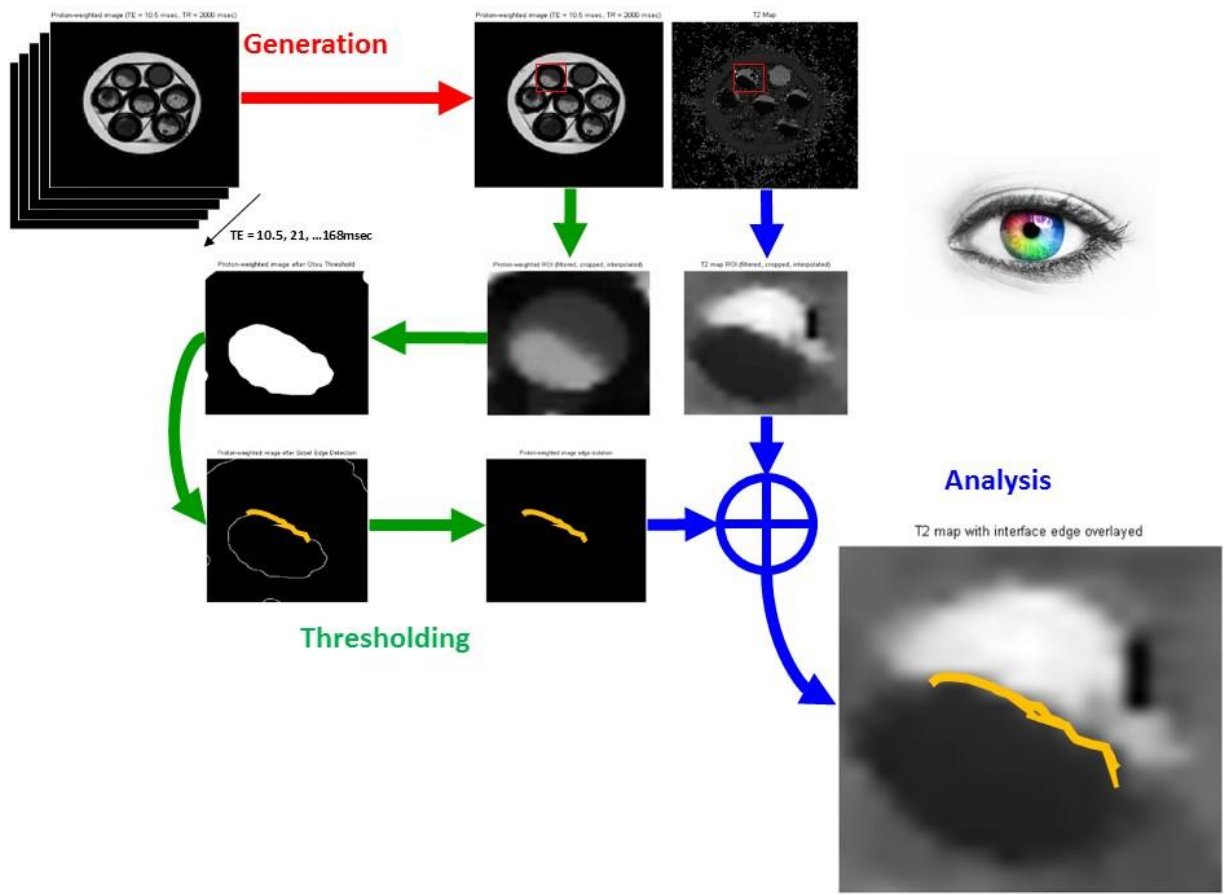

Figure 2. Image analysis begins with the generation of the $T_2$ map from the time-series data. Then, the proton-weighted image undergoes thresholding and edge detection. The pixels in the calculated edge are mapped to the $T_2$ map and averaged.

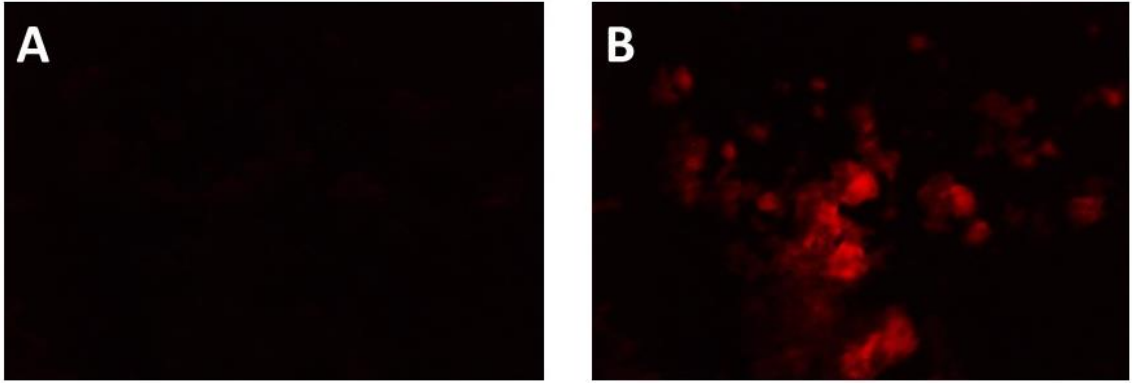

Figure 3. (A) Microscopy image of the UV pre-excited NPs powder with laser stimulation off; (B) Microscopy image of the UV pre-excited NPs powder with laser stimulation on. This demonstrates the stored energy in the pre-excited NPs.

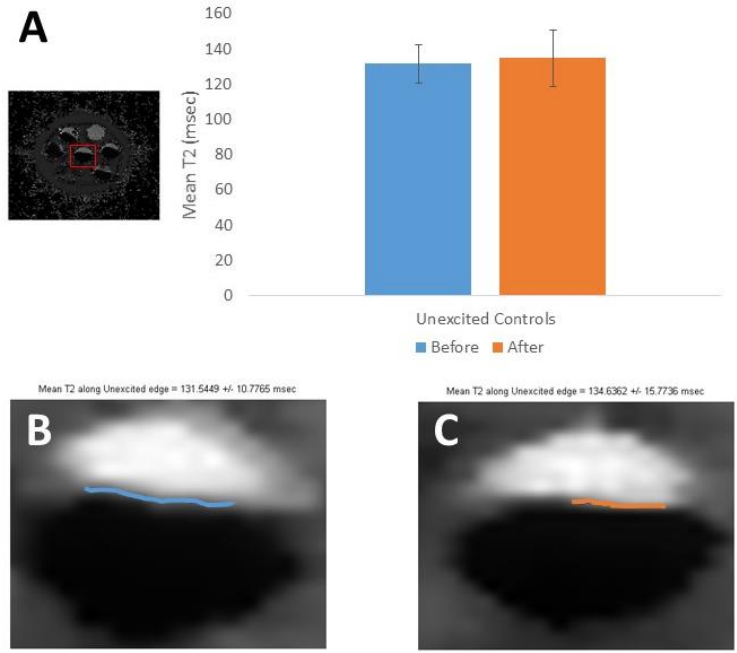

Figure 4. (A) This bar graph demonstrates the average $T_2$ along the mapped edges of the (B) before- and (C) after-excitation images in the unexcited control sample. The mean $T_2$ remains the same along these edge regions as expected.

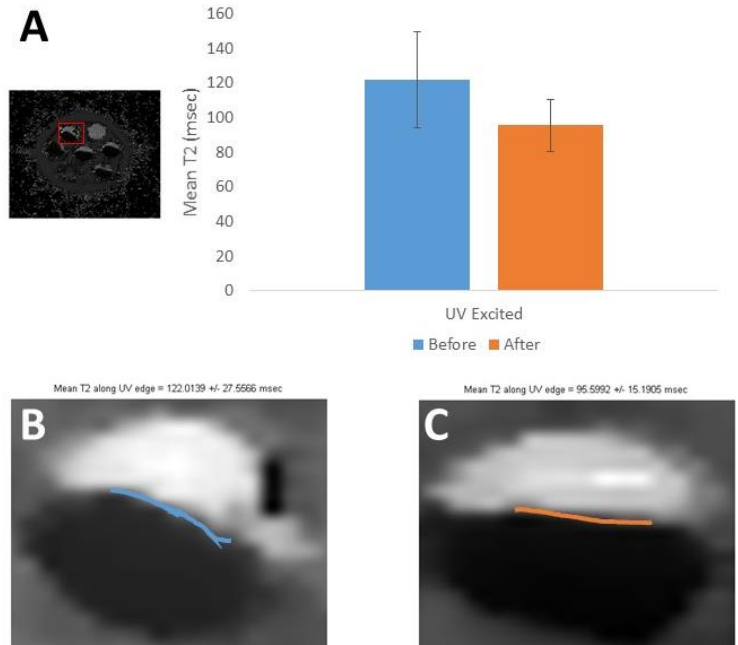

Figure 5. (A) This bar graph demonstrates the average $T_2$ along the mapped edges of the (B) before- and (C) after-excitation images in the UV-excited sample. The mean $T_2$ decreases after the excitation, though the standard deviations do overlap.

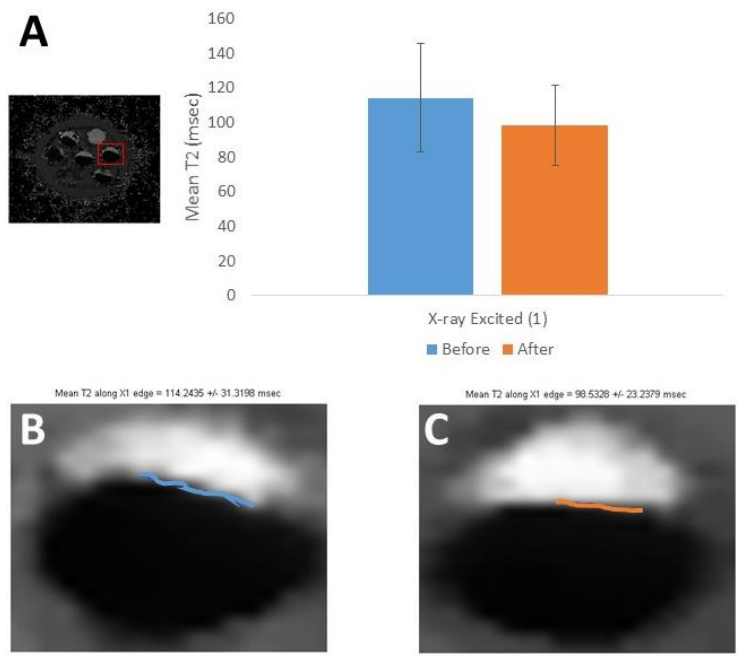

Figure 6. (A) This bar graph demonstrates the average $T_2$ along the mapped edges of the (B) before- and (C) after-excitation images in the first x-ray-excited sample. The mean $T_2$ decreases after the excitation, though the standard deviations do overlap.

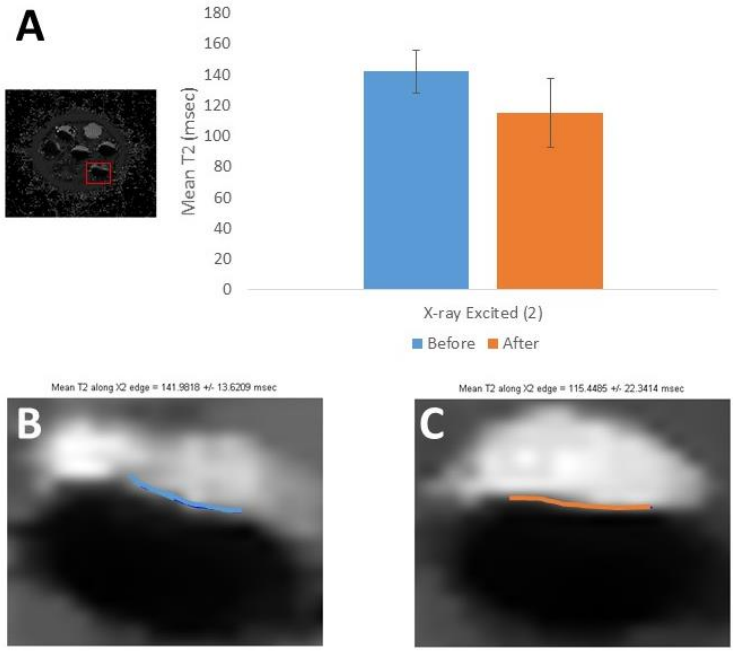

Figure 7. (A) This bar graph demonstrates the average $T_2$ along the mapped edges of the (B) before- and (C) after-excitation images in the second x-ray-excited sample. The mean $T_2$ decreases after the excitation, though the standard deviations do overlap.

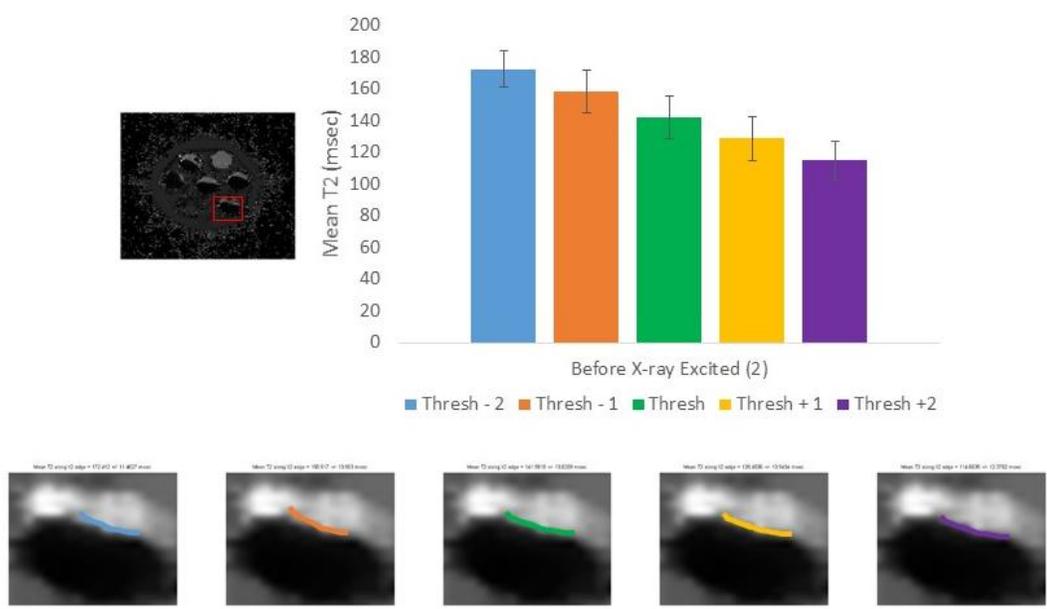

Figure 8. This bar graph demonstrates the change in mean $T_2$ along the second x-ray-excited edge before excitation as the threshold used to calculate the edge is adjusted by +/- 2 grayscale units. Decreasing the threshold results in an increase in mean $T_2$ as it moves the edge further into the water region of the image. Vice versa, increasing the threshold decreases the mean $T_2$. The edges over which the mean $T_2$ is calculated are shown overlaid on the $T_2$ maps along the bottom.

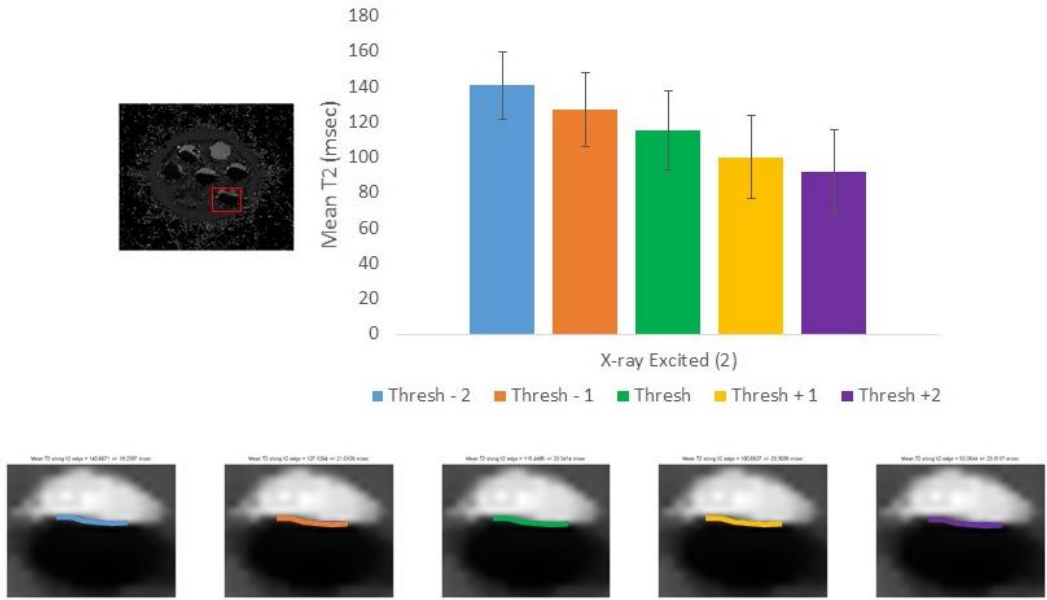

Figure 9. This bar graph demonstrates the change in mean $T_2$ along the second x-ray-excited edge after excitation as the threshold used to calculate the edge is changed by +/- 2 grayscale units. Decreasing the threshold results in an increase in mean $T_2$ as it moves the edge further into the water region of the image. Vice versa, increasing the threshold decreases the mean $T_2$. The edges over which the mean $T_2$ is calculated are shown overlaid on the $T_2$ map. The mean $T_2$ value decreases for each threshold level between the before- and after-excitation images. The edges over which the mean $T_2$ is calculated are shown overlaid on the $T_2$ maps along the bottom.

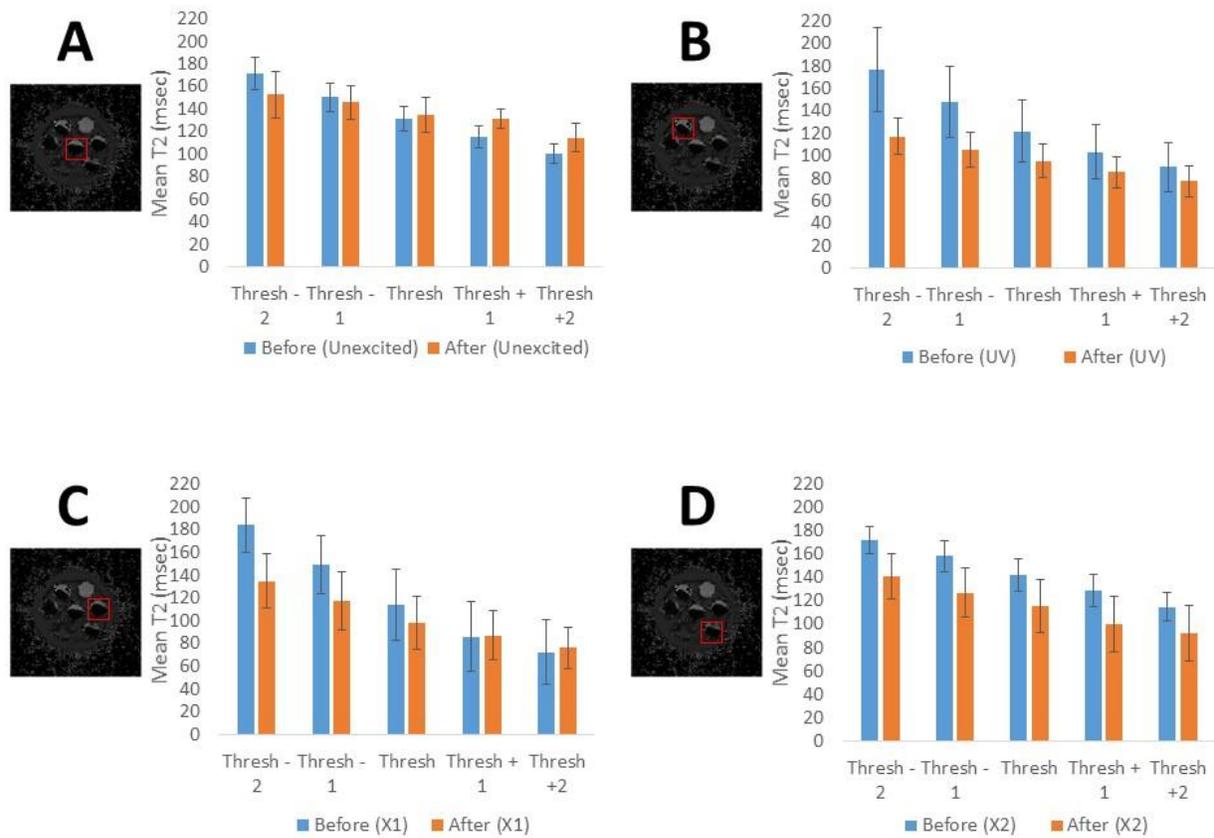

Figure 10. These bar graph compares mean $T_2$s along the calculated edges before- and after- excitation of the (A) unexcited control, (B) UV-excited, (C) first x-ray-excited, and (D) the second x-ray-excited sample. The edges are varied by increasing and decreasing the threshold used to calculate them by 2 grayscale units. Large decreases from before to after-excitation can be seen in each of the excited samples, especially at the edge found deepest into the water region (Thresh – 2).

**Supplemental material:**

Analysis MATLAB Code

```matlab
function [final_T2s,final_T1s] = calculate_edge_relaxation(dcm_T1_image, 
dcm_T2_image, dcm_rho_image, nROIs, ROIs)
% Outputs:
% final_T2s - structure containing fieldnames corresponding to
% those given by input ROIs with values of (1,4)-vector (mean T2, std
% dev ,number of voxels, Otsu threshold level) in edge region of nanoparticle 
deposit.
% final_T1s - structure containing fieldnames corresponding to
% those given input ROIs with values of (1,4)-vector (mean T1, std
% dev ,number of voxels, Otsu threshold level) in edge region of nanoparticle 
deposit.

% Inputs:
% dcm_T1_image - DICOM T1 map as calculated by ParaVision (Bruker machine
% software).  Size should equal dcm_rho_image and dcm_T2_image.
% dcm_T2_image - DICOM T2 map as calculated by ParaVision (Bruker machine
% software).  Size should equal dcm_rho_image and dcm_T1_image.
% dcm_rho_image - DICOM proton density weighted image (short TE, long TR).
% Size should equal dcm_T2_image and dcm_T1_image.
% nROIs - interger representing the number of ROIs to analyze.
% ROIs - cell array of samples names.
% lineLength - length line for perpinduclar profile in pixels.

% Optional outputs:
% final_profiles - structure containing fieldnames corresponding to those
% given by input ROIs and values of intensity profiles along lines normal
% to the edge arranged in arrays.

% Optional inputs:
% lineLength - single or double that sets the length of the profiles

% This program was designed to do in depth analysis of T2 changes after
% nanoparticle excitation via ionizing radiation sources.  The first
% type of particles used in testing were hydrophobic and settled out of
% solution very quickly.  This led to the hypothesis that any changes in
% relaxation as a result of the nanoparticle charging would be seen
% strictly in the region closest to the inteface between water and
% nanoparticle deposit.  The following methods attempt to systematically
% and repeatably find the edge of the nanoparticle deposit and average the
% T2s of the pixels deemed to be the edge.  Some user input is required.

tic

% If the number of samples input does not match the number of sample titles
% given, an error will be returned.
if nROIs ~= length(ROIs)
    error('myApp:argChk', 'ROIs and number of ROIs mismatch')
end

folder = pwd;
mkdir(folder,'post_data');
```

```matlab
% Initializing input arguments and output structures
%lineLength = 30.4*lineLength; % lineLength is multiplied by 30.4 determined
as ratio between interpolated image matrix to initial image matrix
final_T2s = struct([]);
final_T1s = struct([]);
%final_profiles = struct([]);
T1image = 30*(7.62939453125e-3).*double(dicomread(dcm_T1_image));
T2image = 30*(9.5367431640625e-4).*double(dicomread(dcm_T2_image)); %
multiplication factor is specific to dcm scaling as given in visu file
(ParaVision)
rhoimage = (0.001953125).*double(dicomread(dcm_rho_image)); % multiplication
factor is specific to dcm scaling as given in visu file (ParaVision)
srhoimage = medfilt2(rhoimage,[5 5]);
smoothed_image = medfilt2(T2image,[5 5]); % median filters are applied to
both images to denoise
sT1image = medfilt2(T1image,[5 5]);

% Begin loop over ROIs
for ii = 1:nROIs
    % Region selection using T2 map
    z=figure;
    imagesc(smoothed_image);colormap('gray');colorbar
    string = ['Please select region corresponding to ' ROIs{ii}];
    title(string)
    [c_image,rect] = imcrop(z);
    
    % Identical cropping of rho image
    c_rho = imcrop(srhoimage,rect);
    c_T1 = imcrop(sT1image,rect);
    
    % Cubic interpolation of cropped images
    intIs = interp2(c_image,5,'cubic');
    intRhos = interp2(c_rho,5,'cubic');
    intT1s = interp2(c_T1,5,'cubic');
    
    % Further isolate sample
    %     figure
    %     imagesc(intRhos);colormap('gray')
    %     BW = roipoly;
    %     new_intRhos = zeros(size(intRhos));
    %     new_intRhos(BW == 1) = intRhos(BW == 1);
    %     %    imhist(new_intRhos)
    %     close
    
    % Otsu thresholding of rho image to determine nanoparticle/water
    % interface
    level(1) = multithresh(intRhos);
    level(2) = level(1) + 1;
    level(3) = level(2) + 1;
    level(4) = level(1) - 1;
    level(5) = level(4) - 1;
    %     level(2) = level(1)*1.04;
    %     level(2) = level(1)*1.08;
    %     level(4) = level(1)*0.96;
    %     level(3) = level(1)*0.92;
```

```matlab
%         level(6) = level(1)*0.88;
%         level(7) = level(1)*1.12;
%         intIs = interp2(c_image,5,'cubic');
%         intRhos = interp2(c_rho,5,'cubic');
%         levelT = multithresh(inv_T2);
        level = sort(level);
        close(z)
        for i = 1:length(level)
            orhoimage = imquantize(intRhos,level(i));
            
        %      oT2image = imquantize(inv_T2,levelT);
%          close(z)
            
            % Gradient magnitude and angle calculation of rho image
            [Gmag,Gdir] = imgradient(intRhos,'sobel');
            
            % Edge isolation of thresholded rho image
            e_image = zeros(size(orhoimage));
            
            [e_image,thresh] = edge(orhoimage, 'sobel','nothinning');
            
            % Overview of calculated edges
            h=figure;
            subplot(1,2,1)
            imagesc(intIs)
            subplot(1,2,2);
            imagesc(e_image); colormap('gray');
            title('Calculated edges for current ROI. Exit image to continue')
            uiwait(h)
            
            % Isolate connected edges
            CC = bwconncomp(e_image);
            
            % Begin loop over connected edges to select edge at NP/water
interface
            for kk = 1:length(CC.PixelIdxList)
                % Initialize matrix
                new_image = zeros(size(e_image));
                % Create mask image
                matr = cell2mat(CC.PixelIdxList(kk));
                for ll = 1:length(matr)
                    new_image(matr(ll)) = 1;
                end
                % Disconnect unwanted regions
                tt = figure;
                subplot(1,2,1)
                imagesc(intIs); colormap('gray')
                title('Press enter to continue to next edge. If current edge is
the desired, please select two points which define a rectangle in which all
pixels will become zeros (left to right, top to bottom).');
                ttt = subplot(1,2,2);
                imagesc(new_image); colormap('gray')
                title('Multiple rectangles are allowed. Press enter to
continue.')
                discon = ginput;
```

```matlab
        if isempty(discon) == 0
            break
        end
        close
    end
    close
    
    % Make pixels in rectangle equal to zero.
    [row,column] = size(discon);
    for mm = 1:2:row
        for nn = round(discon(mm,2)):round(discon(mm+1,2))
            for oo = round(discon(mm,1)):round(discon(mm+1,1))
                new_image(nn,oo) = 0;
            end
        end
    end
    %new_image = imdilate(new_image,strel('disk',1));
    
    % Display disconnected edges
    h=figure;
    subplot(1,2,1)
    imagesc(intIs); colormap('gray')
    ttt = subplot(1,2,2);
    imagesc(new_image); colormap('gray');
    title('Disconnected edges remaining. Exit image to continue')
    uiwait(h);
    
    % Loop over disconnected regions to select interface edge
    CC2 = bwconncomp(new_image);
    for nn = 1:length(CC2.PixelIdxList)
        new_image2 = zeros(size(new_image));
        matr2 = cell2mat(CC2.PixelIdxList(nn));
        for oo = 1:length(matr2)
            new_image2(matr2(oo)) = 1;
        end
        % Display edge
        figure
        subplot(1,2,1)
        imagesc(intIs); colormap('gray')
        title('Press enter to go to next edge.')
        ttt = subplot(1,2,2);
        imagesc(new_image2); colormap('gray')
        title('Select a point in edge image to select edge.')
        final = ginput;
        if isempty(final) == 0
            break
        end
        close
    end
    close
    number_of_voxels = sum(sum(new_image2));
    %new_image3 = imdilate(new_image2,strel('disk',5,0));
    
    % Isolate edge pixel gradient angles, magnitudes, and T2s
    edgeAngle = zeros(size(new_image2));
    edgeAngle(new_image2 == 1) = Gdir(new_image2 == 1);
```

```matlab
        edgeGmag = zeros(size(new_image2));
        edgeGmag(new_image2 == 1) = Gmag(new_image2 == 1);
        edgeT2s = intIs(new_image2 == 1);
        edgeT1s = intT1s(new_image2 == 1);
        meanedgeT2 = nanmean(edgeT2s);
        edgestd = nanstd(edgeT2s);
        meanedgeT1 = nanmean(edgeT1s);
        edgestdT1 = nanstd(edgeT1s);
        %% Profiles for lines perpindicular to edge
        % Calculate profile lines on interpolated T2 maps
        %       [a_y,a_x] = find(edgeAngle);
        %       Ia_x = a_x;
        %       Ia_y = a_y;
        %       line_x = zeros(1,2*length(Ia_x));
        %       line_y = zeros(1,2*length(Ia_y));
        %       for pp = 0:(length(line_x)/2)-1
        %           line_x((2*pp)+1) = 
Ia_x(pp+1)+lineLength*cosd(edgeAngle(a_y(pp+1),a_x(pp+1)));
        %           line_x((2*pp)+2) = Ia_x(pp+1)-
lineLength*cosd(edgeAngle(a_y(pp+1),a_x(pp+1)));
        %           line_y((2*pp)+1) = Ia_y(pp+1)-
lineLength*sind(edgeAngle(a_y(pp+1),a_x(pp+1)));
        %           line_y((2*pp)+2) = 
Ia_y(pp+1)+lineLength*sind(edgeAngle(a_y(pp+1),a_x(pp+1)));
        %       end
        %
        %       % Display interpolated T2 map
        %       figure
        %       subplot(1,2,1)
        %       imagesc(intIs);colormap('gray');
        %       hc = colorbar;
        %       ylabel(hc, 'T2 (msec)');
        %       hold on
        %
        %       % Plot profile lines on T2 map and compile intensity profiles
        %       for po = 0:(length(line_x)/2)-1;
        %           profiles(po+1,:) = improfile(intIs,[line_x((2*po)+1) 
line_x((2*po)+2)],[line_y((2*po)+1) line_y((2*po)+2)],160);
        %           plot([line_x((2*po)+1) line_x((2*po)+2)],[line_y((2*po)+1) 
line_y((2*po)+2)],'r');
        %       end
        %
        %       % Plot profiles
        %       subplot(1,2,2)
        %       plot(profiles')
        %       hold on
        %       plot(nanmean(profiles)', '-b', 'MarkerSize', 18)
        %
        %       % Compile outputs
        %       f_profiles = struct(ROIs{ii},profiles);
        %       final_profiles = catstruct(final_profiles,f_profiles);

        %% Output organization and figure production
        fieldname = [ROIs{ii} '_' num2str(i)];
        T2s = 
struct(fieldname,[meanedgeT2,edgestd,number_of_voxels,level(i)]);
        final_T2s = catstruct(final_T2s,T2s);
```

```matlab
        T1s = struct(fieldname,
[meanedgeT1,edgestdT1,number_of_voxels,level(i)]);
        final_T1s = catstruct(final_T1s,T1s);

        filename = [folder '\post_data\post_' ROIs{ii} '_T2_' num2str(i)];
        g=figure;
        overlay = imoverlay(mat2gray(intIs),mat2gray(new_image2),[0 0 1]);
        imagesc(overlay)
        string = ['Mean T2 along ' ROIs{ii} ' edge = ' num2str(meanedgeT2) '
+/- ' num2str(edgestd) ' msec'];
        title(string)
        axis off
       % print(g,'-dtiff',filename); % Saves images

        filename = [folder '\post_data\post_' ROIs{ii} '_T1_' num2str(i)];
        figure
        overlay = imoverlay(mat2gray(intT1s),mat2gray(new_image2),[0 0 1]);
        imagesc(overlay)
        string = ['Mean T1 along ' ROIs{ii} ' edge = ' num2str(meanedgeT1) '
+/- ' num2str(edgestdT1) ' msec'];
        title(string)
        axis off
       % print(g,'-dtiff',filename); % Saves images
    end
end

toc
```

Original images are available upon request and will be uploaded to our website: http://www.rpi-bic.org/

# calculate_edge_relaxation.m Tutorial Document

1. **Run calculate_edge_relaxation.m**
   a. [T2_final, T1_final] = calculate_edge_relaxation('..\..\after-excitation_T1_map.dcm', '..\..\after-excitation_T2_map.dcm', '..\..\after-excitation_PD_image.dcm', 4,, {'UV', 'Unexcite', 'X-ray1', 'X-ray2'});
   b. Optional inputs and output are available for further functionalization. See M-file for more information
2. **Crop selected region.**
   a. 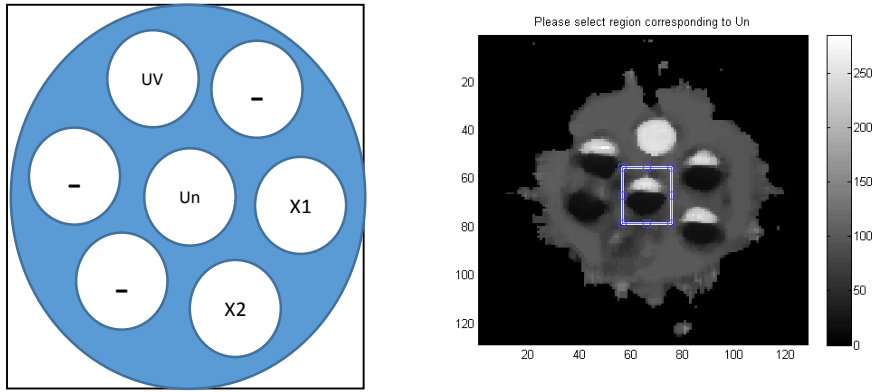

3. **Figure shows $T_2$ map and calculated edges. Exit figure to continue.**

   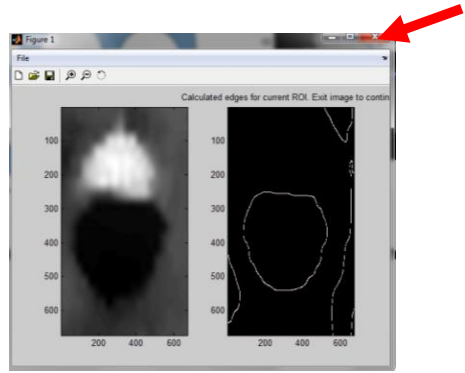

4. **Next series of images isolates each connected edge. Press enter to get to edge of interest. Once at the right edge, select at least two points to make a rectangle to disconnect areas of the edge that are not of importance. Press enter to continue.**

   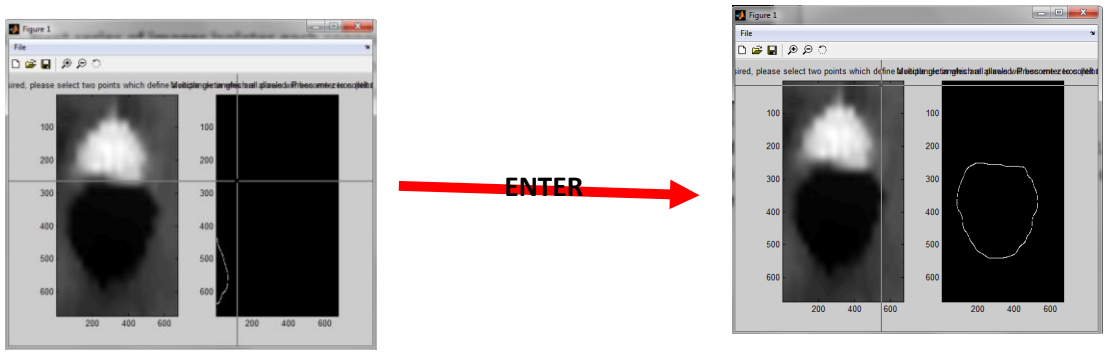

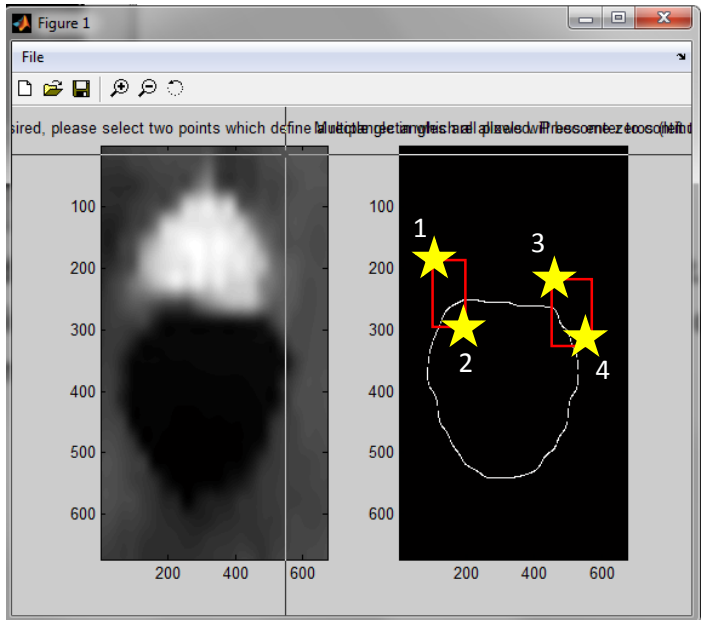

5. Figure shows T$_2$ map and disconnected edges. Exit figure to continue.

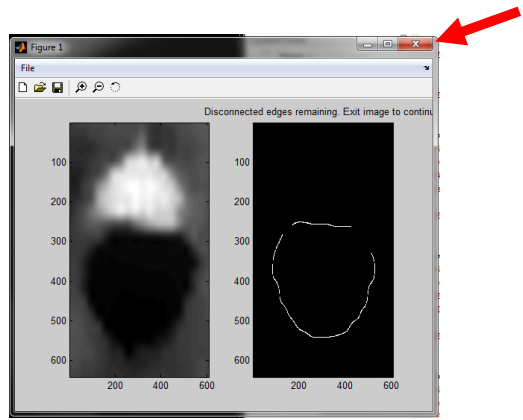

6. Next series of images isolates each new connected edge. Press enter to get to edge of interest. Once at the right edge, select any point in image to select edge. Press enter to continue.

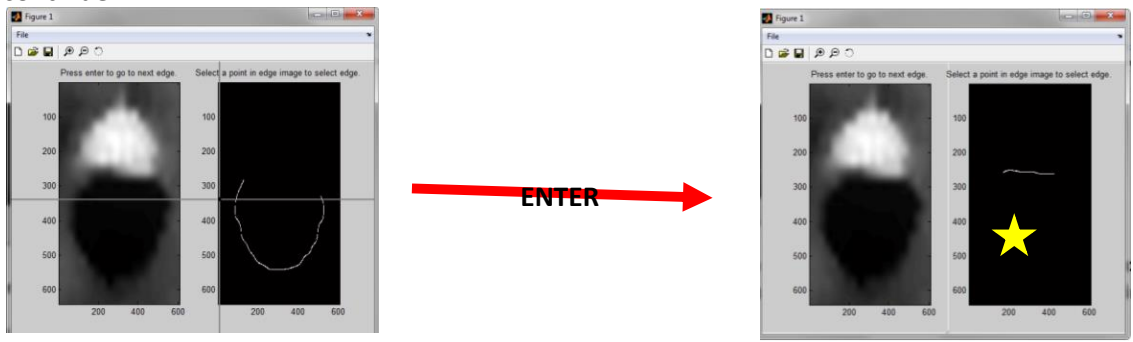

7. **Mean relaxation times are calculated and placed in output structures. See M-file for more info about saving images.**
8. **Repeat 3-7 for each threshold adjustment.**
9. **Repeat 2-8 for each sample.**